# The orientation of Trajan's town of Timgad

**Amelia Carolina Sparavigna**
Department of Applied Science and Technology
Politecnico di Torino, C.so Duca degli Abruzzi 24, Torino, Italy

*An equation, generally used for solar energy applications, can be applied to archaeoastronomical calculations to determine the day of foundation of ancient towns. The equation is that giving the sunrise amplitude, and, the direction of the sunrise on the day of foundation was sometimes used to determine the main street of the town. Here we discuss the case of Timgad, a town founded by the Roman emperor Trajan. Probably, Timgad was founded during the first decade of September.*

Timgad was a Roman colonial town founded by the Emperor Trajan around AD 100. The Roman full name was "Colonia Marciana Ulpia Traiana Thamugadi": in this name we find the names of emperor's mother Marcia, his father Marcus Ulpius Traianus and his eldest sister Ulpia Marciana. The ruins of the Trajan's Timgad are in Algeria. The city was intended to be a defence of the Roman Africa against the Berbers and in origin populated by veterans of Trajan's army. After a long peaceful existence, the city started its decline after being sacked by Vandals in the 5th century. The Arab invasion caused the final ruin of Timgad that ceased to be inhabited after the 8th century [1]. The sands of Sahara covered the city until its excavation in 1881.

It the book entitled "Ancient Town-Planning", written by F. Haverfield and published in 1913 [2], besides Torino, the Julia Augusta Taurinorum, Timgad is proposed as a noteworthy site for being one of the best examples of the Roman city planning. When the author wrote the book, about Trajan's Timgad there were only purely archaeological remains. Haverfield reports that the ruins are on "the northern skirts of Mount Aurès, halfway between Constantine and Biskra and about a hundred miles from the Mediterranean coast. Here the emperor Trajan founded in A.D. 100 a 'colonia' on ground then wholly uninhabited, and peopled it with time-expired soldiers from the Third Legion which garrisoned the neighbouring fortress of Lambaesis ... The 'colonia' of Trajan appears to have been some 29 or 30 acres in extent within the walls and almost square in outline (360 x 390 yds.). It was entered by four principal gates, three of which can still be traced quite clearly, and which stood in the middle of their respective sides; the position of the south gate is doubtful. According to Dr. Barthel, the street which joins the east and west gates was laid out to point to the sunrise of September 18, the birthday of Trajan."

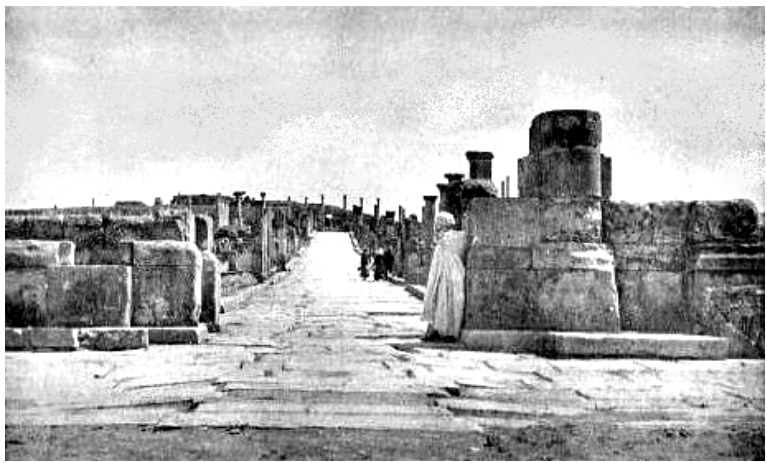

*Streets of Timgad, from a photograph in [2]*

As we will see in the following discussion, it is impossible that it was precisely AD 100 the year of foundation, if we consider that Trajan himself was in Africa for the foundation. But, let us discuss the assertion that the street which joins the east and west gates, that is the decumanus, was laid out to point to the sunrise of September 18.

In [3], it is told how sometimes the Romans, and before Greeks and Etruscans, used to orient their towns with the sunrise on the day of the foundation. The ancient Julia Augusta Taurinorum, has the main street probably oriented according to a sunrise on the winter solstice or in January. After this main direction and its perpendicular had been determined, the rectangular area of the new town was subdivided in a chessboard of "insulae", the modern house-blocks [4].

Let us consider the Trajan's town of Timgad as we can see in the satellite images: we note that the main street, the decumanus, possesses an angle of 5.5 degrees with respect to the cardinal East-West direction (see Fig.1). It is then impossible that the town was founded on September 18, because this day is close to the autumn equinox, and on equinoxes the sun is rising due East.

To see on what day Timgad was founded, we can use the sunrise amplitude equation, as we did for Torino [3]. This is an equation derived for solar energy application and providing the angle the direction of the rising sun is forming with the cardinal East-West direction.

We need to know the latitude and the declination. In [5-7] it had been proposed and used a formula for the declination as a function of the days after the spring equinox, as reported in the following table. The declination, with the hour angle ω, are the coordinates in the equatorial system. To have the sunrise amplitude Z, that is the direction of the sun on the observer's horizon, we use the last equation in the table, as that given in http://www.titulosnauticos.net/astro/

latitude $\varphi$

declination (in radians)

$$\delta = arcsin(0.4 \cdot sin(2\pi n / 365))$$

$n$ = number of days after the spring equinox

hour angle $\omega$

$$\omega = \frac{360°}{2\pi} arccos(-tan\varphi \cdot tan\delta) - 90°$$

sunrise amplitude

$$Z = 90° - \frac{360°}{2\pi} arccos(sin\delta / cos\varphi)$$

A quite useful visual representation of the direction of sunset, noon, and sunrise for a location is provided at the site http://www.sollumis.com/. This is implemented on Google Maps, suitable for any location on the map.

Hour angle and amplitude have different values. If we decide to use the hour angles plotted as a function of days, we guess that who founded the town used an equatorial reference. In the case that we use the sunrise amplitude we consider the foundation according to the horizontal coordinate system.

Let us therefore plot the hour angle and the amplitude, on the Timgad latitude, as a function of the days after the spring equinox.

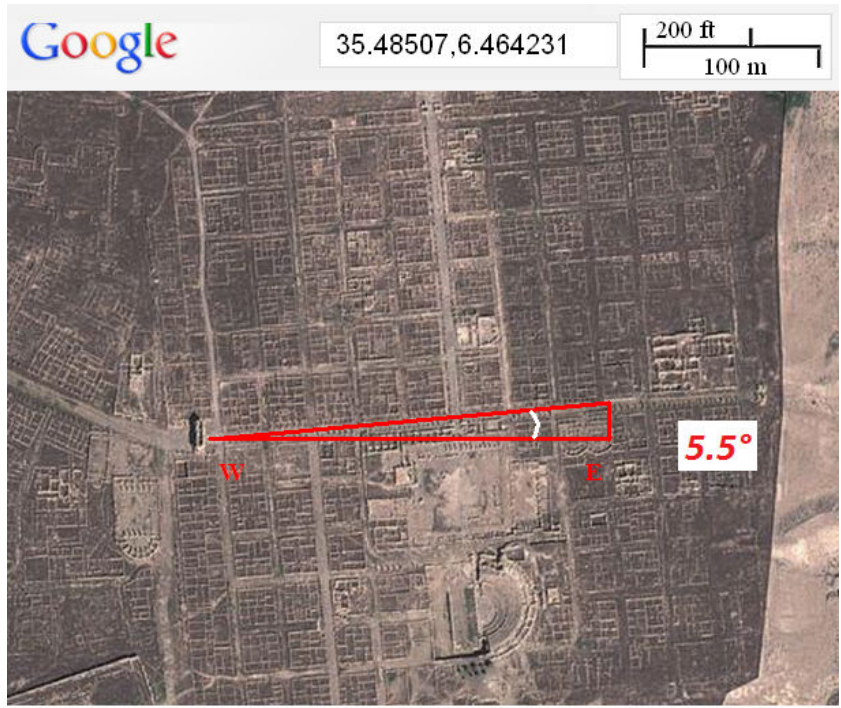

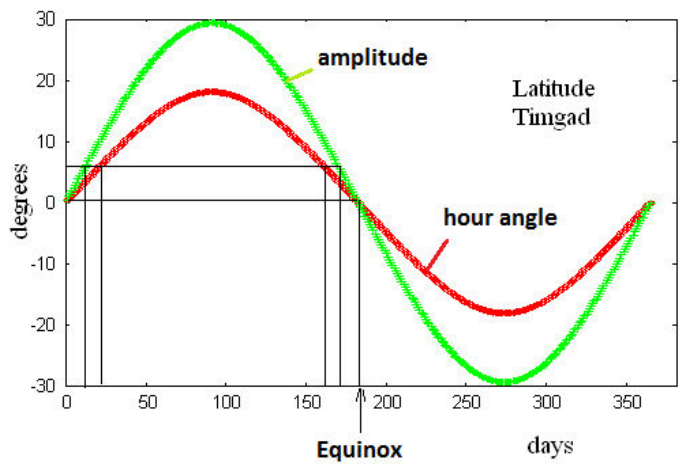

*Fig.1. In the upper part of the figure we see the satellite map of Timgad. Using two sides of the right-angled triangle (one is the East-West direction) we find the angle. In this case it is of 5.5° (positive). The hypotenuse is the decumanus, the main street. In the lower part, we see the sunrise amplitude (green) as a function of the days after the spring equinox, evaluated at Timgad latitude. Let us note that there are two possible days corresponding to an angle of 5.5°, one is approximately 10 days after the spring equinox and the other approximately 10 days before the autumn equinox For comparison, the hour angle (red) is also shown, in this case the angle of Timgad corresponds to 20 days after the spring equinox and 20 days before the autumn equinox.*

Let us suppose that, at foundation, it was decided to align the decumanus with the direction of the rising sun. It means that measuring the angle of the decumanus in satellite images, we can see on what day after the spring equinox the orientation of Timgad is corresponding. From Fig.1, we have that the town could had been founded approximately 10 days after the spring solstice or approximately 10 days before the autumn equinox. If we choose the hour angle, we have the that Timgad could had been founded approximately 20 days after the spring solstice or approximately 20 days before the autumn equinox

We can ask ourselves what is the correct solution. To answer we need some information on Trajan's life and on Roman traditions in general.

Trajan (Marcus Ulpius Nerva Traianus Augustus) was born on 18 September 53 CE in the province of Hispania Baetica, Trajan served Rome in Hispania Tarraconensis and was a supporter of the emperor Domitian. In September 96, Domitian was succeeded by Marcus Cocceius Nerva, an old and childless senator. Nerva named Trajan as his adoptive son and successor. It seems that on October 97, he ascended the Capitol and proclaimed that he adopted Trajan as his son. The senate confirmed the choice and acknowledged the emperor's adopted son as his successor. When Nerva died on 27 January 98, Trajan succeeded without incident [8].

And this is enough for our discussion on the day of the foundation of Trajan's Timgad. September 18 is usually designated as the Trajan's birthday from a letter written by Pliny the Younger. This was the day of Domitian's death and when Nerva was raised to the empire [9]. It is possible that it was not the true birthday of Trajan, but that of his adoption, or the anniversary of his accession, but in any case, these days do not seems to be coincident with the day of the foundation of Timgad, as we can propose using the amplitude or the hour angle.

If we consider the amplitude, we could guess that Trajan decided to found Timgad around 10 September, a period that Rome devoted to the festival of "Ludi romani" [10], sacred to Jupiter. We can propose this possibility after the use of the sunrise amplitude equation and the satellite images of the site.

If we consider the hour angle, there is another day, concerning Trajan's life, which could be in agreement with the orientation of Timgad and this is the first day of September. On that day, the year 100 CE, Pliny the Younger was appointed to deliver his "Panegyric", facing Trajan and the Senate [11,12]. In that year therefore, Trajan was in Rome. The first day of the month was the Kalends, "Kalendae" [13,14], and the Kalends of September was a "dies fasti", on which legal actions are permitted [15]. And this is why Pliny made his speech on that day in the Senate, because he was starting his consulate.

From this episode of Trajan's life we learn that the Kalends were traditionally a good day and therefore suitable for the foundation of a new town. It is then quite possible that, around 100 CE, Trajan decided to found Timgad on the Kalends of September, sacred to Juno the patron goddess of Rome and the Roman Empire. In any case, it seems that probably the town was founded during the first decade of September.

Editor, Rue des Mathurins St. Jacques n.5, Paris, MDCCCXLV, In a note at Pag.241, etitled "Quem tuus natalis exornat", Burnouf is telling the following "L'ensemble de cette phrase, et la Lettre 23 du liv. X, prouvent que Pline désigne le 18 septembre, jour où périt Domitien (18 September 96, Suét., 17), et où Nerva fut élevé à l'empire. Trajan était-il aussi né ce jour-là? Cette coïncidence n'est pas impossible, mais elle est au moins fort, remarquable. Aussi Schwartz essaye-t-il de prouver que ce n'est pas de la naissance de Trajan qu'il s'agit, mais de son adoption. Les mots natalis dies peuvent évidemment s'employer dans ce sens, et natalis imperii peut signifier même l'anniversaire de l'avénement d'un prince: les exemples cités par Schwartz ne laissent aucun doute sur ce point, non plus que sur le sens figuré que "genuit" est susceptible de recevoir. Cependant les arguments de ce critique, à défaut de preuves directes, devraient établir deux impossibilités, l'une queTrajan soit né le 18 septembre, l'autre qu'il ait été adopté un autre jour que le 18 septembre, et c'est ce qu'ils ne font pas. Tillemont discute la même question, et, sans rien prononcer, il ne nie pas que natalis ne puisse s'entendre du jour de l'adoption. J'ai traduit les mots latins dans leur sens le plus naturel, me réservant d'avenir le lecteur d'une difficulté qui a partagé les plus habiles critiques." This note is telling that the Letter shows that Pliny designates September 18, the day of the death of Domitian, and when Nerva was raised to the empire. Trajan was also born there that day? This coincidence is not impossible, but it is at least very remarkable. Schwartz tries to prove that this is not the birthday of Trajan, but its adoption. The words "dies natalis" can obviously be used in this sense, and "imperii natalis" can even be the anniversary of the accession. Burnouf therefore translated the Latin words in their most natural sense.

[10] http://it.wikipedia.org/wiki/Festività_romane#September
[11] A panegyric is a public speech, delivered in high praise of a person. It is derived from the Greek, meaning "a speech fit for a general assembly". The Romans generally confined the panegyric to the living, and reserved the funeral oration for the dead. The most celebrated example of a Latin panegyric is that delivered by the Pliny the Younger. This "Panegyric" is most important source of information on the emperor.
[12] Mary Whitby, The Propaganda of Power: The Role of Panegyric in Late Antiquity, BRILL, 1998
[13] http://en.wikipedia.org/wiki/Roman_calendar
[14] http://penelope.uchicago.edu/~grout/encyclopaedia_romana/calendar/juliancalendar.html
[15] http://thepaganleft.blogspot.it/2005/09/kalends-of-september.html